\documentclass[12pt, a4paper]{article}

\usepackage[latin1]{inputenc}
\usepackage[T1]{fontenc}

\usepackage{amsmath}
\usepackage{amssymb}
\usepackage{amsfonts}
\usepackage{amsthm} 
\usepackage{bbm}
\usepackage{mathtools}

\usepackage{graphicx}
\usepackage{float}
\usepackage{caption}
\usepackage{diagbox}
\usepackage{longtable}
\usepackage{multirow}
\usepackage{booktabs}

\usepackage{hyperref}
\usepackage{biblatex}
\usepackage{xcolor}
\usepackage{multicol}
\usepackage{enumitem}
\usepackage{bm}
\usepackage{algorithm}
\usepackage{algpseudocode}

\usepackage{tikz}
\usepackage{comment}

\newcommand{\ssitem}[1][black]{\stepcounter{enumii}\item[\color{1}$\bm{*}$\,\textbf{(\alph{enumii})}]}
\newcommand{\sitem}[1][black]{\stepcounter{enumi}\item[\color{1}$\bm{*}$\,\theenumi.]}
\newcommand{\dsitem}[1][black]{\stepcounter{enumi}\item[\color{1}$\bm{**}$\,\theenumi.]}

\setlist[enumerate,1]{label=\textbf{\arabic*.}}
\setlist[enumerate,2]{label=\textbf{\alph*)}}

\newtheoremstyle{noparens}%
    {}{}                  
    {\itshape}{}          
    {\bfseries}{}         
    { }                   
    {\thmnote{}}        

\theoremstyle{noparens}

\newtheorem{definition}{Definition}[section] 

\theoremstyle{noparens}

\title{
A Hype-Adjusted Probability Measure \\
for NLP Stock Return Forecasting}
\author{
    Zheng Cao\thanks{Graduate Student, Johns Hopkins University} \\
    \texttt{zcao26@jh.edu} 
    \and 
    H\'{e}lyette Geman\thanks{Research Professor, Johns Hopkins University} \\
    \texttt{hgeman1@jhu.edu}
}

\date{}

\begin{document}

\maketitle

\begin{abstract}

This article introduces a Hype-Adjusted Probability Measure in the context of a new Natural Language Processing (NLP) approach for stock return and volatility forecasting. A novel sentiment score equation is proposed to represent the impact of intraday news on forecasting next-period stock return and volatility for selected U.S. semiconductor tickers, a very vibrant industry sector. This work improves the forecast accuracy by addressing news bias, memory, and weight, and incorporating shifts in sentiment direction. More importantly, it extends the use of the remarkable tool of change of Probability Measure developed in the finance of Asset Pricing to NLP forecasting by constructing a Hype-Adjusted Probability Measure, obtained from a redistribution of the weights in the probability space, meant to correct for excessive or insufficient news.

\end{abstract}


\newpage
\section{Introduction} \label{Section: Introduction}

Sentiment has long been recognized as a key driver in financial markets, influencing both trading behavior and asset prices. Previous studies by Hirshleifer (2009)  
have examined the theoretical and empirical role of sentiment in market dynamics, highlighting its impact on investors' decision-making and trading patterns\cite{hirshleifer2009}. Furthermore, research \cite{baker2006investor} by Baker and Wurgler (2006) followed by others has shown a significant relationship between investor sentiment and stock returns, providing a foundation for sentiment-based trading strategies.

The NLP part of our manuscript draws inspiration from the work of Deveikyte et al (2022) \cite{deveikyte2022}, who applied Latent Dirichlet Allocation to forecast market prices and volatility. The authors proposed a thorough approach to compute the sentiment score and forecast the directions of the next day's market return and volatility in the case of FTSE100 stocks. In this research, we choose semiconductors, a very vibrant industry of this decade, as the sector for analyzing news and market data. 

After a review of the literature on NLP and Changes of Measure in section \ref{Section: Literature Review}, a detailed overview of data and methodology is presented in section \ref{Section: Data and Methods}, which includes the developments from previous sentiment equation to more sophisticated ones along with enhanced NLP forecasting methods. Building upon these foundations, we introduce in section \ref{section: Hype-Adjusted Probability Measure}, a novel probability measure, that we name `` Hype-Adjusted Probability Measure", designed to capture the occurrence of market ``hype." The results and future work are discussed in section \ref{section: Model Results}, the forecasting accuracy is significantly increased by $8 \%$.

To our best knowledge, no paper has yet introduced a change of probability measure in the context of NLP forecasting of asset returns and volatility. The instrument is beautiful, the application here original. 

\newpage
\section{Literature Review}
\label{Section: Literature Review}

Sentiment analysis, a subfield of Natural Language Processing (NLP), focuses on quantifying the emotional tone and intent conveyed in textual data. NLP itself is a rapidly growing area of machine learning (ML) that enables computers to process and understand human language through algorithms and statistical models such as the founding work by Jurafsky and Martin (2000) \cite{jurafsky2009speech}. Within NLP, sentiment analysis applies techniques to assess positive, negative, or neutral sentiment in data, often used for applications like market forecasting and consumer behavior analysis \cite{liu2012sentiment}.

In this context, sentiment is defined as the emotional polarity associated with a piece of text, often derived from news articles, tweets, or other media. It is typically measured using computational methods such as dictionary-based approaches or ML models. For example, Vader, a lexicon-based sentiment analysis tool, evaluates sentiment scores ranging from extremely negative to extremely positive \cite{hutto2014vader}.

In their 2022 paper, Deveikyte et al identified a significant negative correlation between daily negative sentiment and observed market volatility and develop a sentiment-based model for predicting directional volatility and market returns \cite{deveikyte2022}. Building on their findings, we aimed to enhance the predictive relationship between news sentiment and stock performance by refining the methods used to calculate key financial metrics.

The return on day \( t \) are classically defined as the logarithmic change in the closing price from day \( t-1 \), expressed as:

\begin{equation}
r_t = \log \frac{\text{CLOSE}_t}{\text{CLOSE}_{t-1}}
\end{equation}

The annualized volatility,  \( \sigma \), is calculated by the formula:

\begin{equation}
\sigma = \sqrt{\frac{1}{N} \sum_{t+1}^{N} (r_t - \bar{r})^2 } \cdot \sqrt{252}
\end{equation}

Notice, for the programming and ML components of the manuscript, we selected a rolling window of 5 days to account for the amount of a regular trading week (without holiday breaks).

These metrics serve as the foundation for analyzing the connection between sentiment data and stock market behavior.

We adopted the Vader Sentiment engine to help translate multi-language texts into scores. It classifies sentiment scores on a scale as introduced by Clayton J. Hutto and Eric Gilbert \cite{hutto2014vader}:
\begin{enumerate}
    \item Extremely negative: -4
    \item Neutral: 0
    \item Extremely positive: +4
\end{enumerate}

The sentiment score has been defined by many authors, in particular the paper \cite{gabrovsek2016twitter} as the fraction of the difference between the number $N$ of positive and negative sentiment tokens, scaled by the sum of positive, neutral, and adjusted negative cases, of a given time $d$:

\begin{equation}
    Sent_d = \frac{N_d(\text{pos}) - N_d(\text{neg})}{N_d(\text{pos}) + N_d(\text{neutral}) - N_d(\text{neg}) + 3}
\end{equation} \label{equation 3}

Glasserman and Mamaysky (2017)  highlight the predictive power of unusual news patterns in forecasting market stress \cite{glasserman2017unusualnews}. Shapiro et al (2022) developed a novel sentiment-scoring model tailored to economic news articles, showing that daily news sentiment predicts consumer sentiment movements and macroeconomic responses to sentiment shocks, such as increased consumption, output, and interest rates \cite{shapiro2020measuring}. Cohen et al. (2023) demonstrate the robustness of multimodal classifiers under perturbation attacks and present the potential for integrating resilient, multimodal approaches into financial sentiment analysis for improved market forecasting \cite{cohen2023masking}.

\label{Section: Literature Review for Change of Measure}

The change of probability measures has been revealed as a remarkable tool in the finance of Asset Pricing. It started with Paul Samuelson's ``Proof that properly anticipated prices fluctuate randomly" (1965) that he identified the importance of change of probability measure in Finance and prices of futures contracts are martingales \cite{samuelson1965}. Harrison and Kreps (1979) proposed under No Arbitrage and constant interest rates a new risk-neutral (equivalent martingale) probability measure that incorporates the risk premium attached to equities and leads to martingales for discounted stock prices\cite{HarrisonKrep1979}.  The  ``forward measure" introduced in Geman (1989) \cite{Geman1989} was constructed to address the challenges created by stochastic interest rates in the valuation of risky cash flows.

To recall, given a random variable \( X \) and a filtration \( \mathcal{F} \), the conditional expectation \( \mathbb{E}[X \mid \mathcal{F}] \) is defined as:

\begin{equation}
    \mathbb{E}[X \mid \mathcal{F}] = \int_{\Omega} X(\omega) \, d\mathbb{P}(\omega \mid \mathcal{F}),
    \label{eq:conditional_expectation}
\end{equation}

where \( \mathcal{F} \) represents the information up to the current time, and \( \Omega \) is the set of states of nature. This concept underpins the mathematical framework of the Radon-Nikodym derivative, used to construct new probability measures like the risk-neutral or hype-adjusted measures in financial applications.


Note that in the context of product management, Wind and Mahajan (1987) emphasized the value of ``marketing hype" when launching new products in order to create a supportive market environment.


\newpage
\section{Data and Methods}
\label{Section: Data and Methods}

As in the foundational paper of Bengio et al (2003) \cite{bengio2003neural}, we define in this paper `news' as textual data, extracted  from verified financial sources reporting on market trends, corporate developments, and stock performance. Primary data sources for this study were obtained from LSEG and the Eikon API \cite{LSEG}, which provide access to over half a million news articles related to 30 semiconductor stocks. These data sources offer a comprehensive coverage of market events, corporate news, and sentiment indicators.

To mitigate potential biases in the collected data, particularly those arising from media overrepresentation or underrepresentation of specific events, we implemented two primary adjustments:

\begin{enumerate}
    \item Assigning Weights to News Sources: Each news source is assigned a weight to achieve a more objective sentiment score adjustment and the adjusted  score is calculated as:
    \begin{equation}
        \text{sentiment}_{\text{adj}} = \alpha \cdot \text{sentiment} + \beta
    \end{equation}
    where \( \alpha \) represents the weighting factor applied to a news source, and \( \beta \) corrects for inherent biases.

    \item Adjusting for Over- or Under-Reported Events: To address the uneven representation of events in the dataset, we use two key methods:
    \begin{itemize}
        \item Removal of duplicate or near-identical articles to prevent overrepresentation of certain events.
        \item Increase of the impact of under-reported events by assigning higher weights explained below.
    \end{itemize}
\end{enumerate}

\subsection{Data and LDA Model Results}

Appendix \ref{app:ticker_weight_table} presents a sample table of 30 tickers selected from the iShares Semiconductor ETF (SOXX). Note this ticker weight table is adjusted by removing 2 values: CME E-MINI S\&P500-technology sector index future for September 2024, $0.15247 \%$ and CME Index and Options Market E-MINI Russell 2000 index future for September 2024, $0.15247 \%$.  The reason is that these two entries do not represent any actual company, but a mix of companies that are already in our list.

We first assemble the results using the previous sentiment methods on the semiconductor data:

\begin{enumerate}
    \item the ``daily weighted average" sentiment score: computed by the aggregated news data based on individual tickers, and adjust the processed sentiment score by individual tickers' weight from the overall SOXX sector (excluding the 2 Future ETFs removed)
    \item the ``overall daily average" sentiment score: searched by ticker name, without being adjusted by individual weights
    \item the ``overall daily average semi title" sentiment score: searched by the topic of SOXX sector, without being adjusted by individual weights
\end{enumerate}

Each data set generates a accuracy precision report through a simple Linear Discriminant Analysis algorithm \cite{balakrishnama1998lda}.  Linear Discriminant Analysis (LDA) is a commonly used dimensionality reduction technique that is effective for classification. It projects data onto a lower-dimensional space while maximizing separation between classes by modeling inter-class differences using the mean and variance within each class. Unlike Principal Component Analysis (PCA), which maximizes variance, LDA prioritizes class separability, making it especially useful in supervised learning with labeled data.

Tables \ref{tab:classification_report_1},  \ref{tab:classification_report_2}, and \ref{tab:classification_report_3} below present the forecast results based on the simple LDA model, with volatility direction as the predicted target and sentiment scores as the only input training parameters.

For the ``daily weighted average" sentiment score:

\begin{table}[H]
\centering
\begin{tabular}{lcccc}
\hline
\textbf{Class} & \textbf{Precision} & \textbf{Recall} & \textbf{F1-Score} & \textbf{Support} \\
\hline
0 & 0.65 & 0.73 & 0.69 & 33 \\
1 & 0.62 & 0.54 & 0.58 & 28 \\
\hline
\textbf{Accuracy} & \multicolumn{4}{c}{0.64} \\
\textbf{Macro Avg} & 0.64 & 0.63 & 0.63 & 61 \\
\textbf{Weighted Avg} & 0.64 & 0.64 & 0.64 & 61 \\
\hline
\end{tabular}
\caption{Classification Report for Best Volatility Direction Model}
\label{tab:classification_report_1}
\end{table}

For the second, the ``overall daily average" sentiment score:

\begin{table}[H]
\centering
\begin{tabular}{lcccc}
\hline
\textbf{Class} & \textbf{Precision} & \textbf{Recall} & \textbf{F1-Score} & \textbf{Support} \\
\hline
0 & 0.58 & 0.91 & 0.71 & 32 \\
1 & 0.73 & 0.28 & 0.40 & 29 \\
\hline
\textbf{Accuracy} & \multicolumn{4}{c}{0.61} \\
\textbf{Macro Avg} & 0.65 & 0.59 & 0.55 & 61 \\
\textbf{Weighted Avg} & 0.65 & 0.61 & 0.56 & 61 \\
\hline
\end{tabular}
\caption{Classification Report for Best Volatility Direction Model}
\label{tab:classification_report_2}
\end{table}

For the third data input, the ``overall daily average title" sentiment score, the result is 

\begin{table}[H]
\centering
\begin{tabular}{lcccc}
\hline
\textbf{Class} & \textbf{Precision} & \textbf{Recall} & \textbf{F1-Score} & \textbf{Support} \\
\hline
0 & 0.66 & 0.87 & 0.75 & 31 \\
1 & 0.80 & 0.53 & 0.64 & 30 \\
\hline
\textbf{Accuracy} & \multicolumn{4}{c}{0.70} \\
\textbf{Macro Avg} & 0.73 & 0.70 & 0.70 & 61 \\
\textbf{Weighted Avg} & 0.73 & 0.70 & 0.70 & 61 \\
\hline
\end{tabular}
\caption{Classification Report for Best Volatility Direction Model}
\label{tab:classification_report_3}
\end{table}

The latter sections present a more sophisticated method of processing the sentiment score, with supporting results provided.

\begin{figure}[H]
    \centering
    \includegraphics[width=0.75\textwidth]{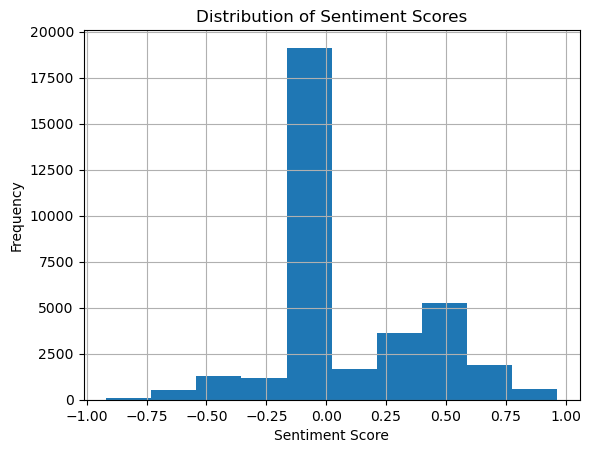}
    \caption{Distribution of Sentiment Scores}
    \label{fig:Distribution_semtiment}
\end{figure}

Figure \ref{fig:Distribution_semtiment} presents a sample distribution of calculated sentiment scores, showing a notable negative skew.

This serves as a motivation for adjusting bias weights, as discussed in Section \ref{section: Bias, Memory, and Weights}, which account for market `hypes' and form the basis for the proposed hype-adjusted probability measure. Bias weights are adjusted to correct for distortions such as excessive coverage of certain stocks and underrepresentation of others. These adjustments ensure a more objective and balanced assessment of market dynamics. Without such corrections, the analysis may disproportionately emphasize certain stocks or news sources, resulting in skewed forecasts or inaccurate conclusions.

\subsection{Bias, Memory, and Weights}
\label{section: Bias, Memory, and Weights}

This section explores the foundational elements - bias, memory, and weights - that underpin the construction of a refined sentiment equation, serving as the basis for the proposed hype-adjusted probability measure. The goal is to systematically address and correct biases arising from media attention imbalances and market weighting discrepancies while incorporating the dynamic influence of past sentiment.

One significant limitation of the prior sentiment equation 3 is that it only counts the number of positive, negative, and neutral news. A news of sentiment score $-3$ should carry a bigger weight than a $-0.1$ score.

We propose a modified approach to sentiment scoring. First, an interval of $(-0.05, +0.05)$ is chosen to broaden the range of sentiment scores classified as neutral, setting all scores within this domain to $0$. Additionally, sentiment scores of extreme values, such as +4, are given higher weights than minimal positive values (e.g., +0.1), addressing the limitation in the original model where they had equal weights.

\begin{equation}
    Sent_d = \frac{1}{n} \sum_{i=1}^n SentScore_{i, d}
\end{equation}

\paragraph{Notations}
\begin{enumerate}
    \item[$1.$] $n$: The total number of tickers in the portfolio.
    \item[$2.$] $Sent_d$: The compound daily sentiment score of an underlying asset portfolio, calculated without considering historical data.
    \item[$3.$] $SentAll_d$: The compound daily sentiment score incorporating the cumulative influence of historical data.
    \item[$4.$] $SentScore_{i,d}$: The average sentiment score for ticker $i$ on date $d$, calculated from a total of $k$ news articles' sentiments $S_{i,d,j}$.
\end{enumerate}
\label{list_notations}

\begin{equation}
    \begin{aligned}
        SentScore_{i, d } =  \frac{1}{k} \ \sum_{j=1}^k  S_{i,d,j} \,  
    \end{aligned}
    \label{eq:SentScore}
\end{equation}

\subsubsection{News Bias}

News bias refers to the disparity in media coverage that individual stocks or sectors receive, which may not align with their actual market significance. For instance, major companies like Nvidia tend to dominate media attention, potentially distorting the sentiment analysis for the entire sector.

The ticker news count weight represents the proportion of total news articles attributed to a specific ticker compared to all tickers within the sector. This metric quantifies the relative attention a ticker receives from the media and audience in comparison to other tickers in the same domain or sector.  Formally, for ticker \( i \):

\begin{equation}
\text{Ticker news count weight}_i = \frac{\text{News count}_i}{\sum_{j=1}^{n} \text{News count}_j},
\end{equation}

where \( n \) is the total number of tickers in the sector. This weight reflects how much media coverage a particular ticker receives relative to others.

The new bias arises from the gap between the Ticker news count weight and the actual market weight of a ticker. Mathematically:

\begin{equation}
\text{Bias}_i = \text{Ticker news count weight}_i - \text{Market weight}_i,
\end{equation}

where:
\begin{itemize}
    \item A positive bias (\( \text{Bias}_i > 0 \)) indicates over-representation in the news.
    \item A negative bias (\( \text{Bias}_i < 0 \)) suggests under-representation in the news.
\end{itemize}

Adjusting for this bias is essential to ensure that the sentiment analysis accurately reflects market dynamics rather than being skewed by excessive coverage or neglect of specific tickers. Without such adjustments:
\begin{itemize}
    \item Over-represented tickers (e.g., Nvidia) may disproportionately influence sector sentiment, leading to inflated predictions.
    \item Under-represented tickers may not properly contribute to the overall sentiment.
\end{itemize}

By integrating both Ticker news count weight and Market weight into the sentiment equation, we aim to balance these disparities and improve the accuracy of market sentiment models.

We examine news bias within the iShares Semiconductor ETF (SOXX) of the U.S. stock market by analyzing how individual asset components contribute to the overall index. Several major companies, such as Nvidia, dominate the sector, with Nvidia representing over $8\%$ of the market. However, over the previous 15 months (up to mid-July 2024), Nvidia had received disproportionately high news coverage of $24.52\%$ compared to other companies in the sector. Considering 30 companies from the SOXX, we assigned weights to each ticker based on their market share. For example, Nvidia (NVDA), valued at \$1,120,866.62 million and accounting for 8.64\% of the sector, has a component weight of 8.64\%. Please refer to Appendix \ref{app:ticker_weight_table} for more details.

\begin{figure}[H]
    \centering
    \includegraphics[width=0.8\textwidth]{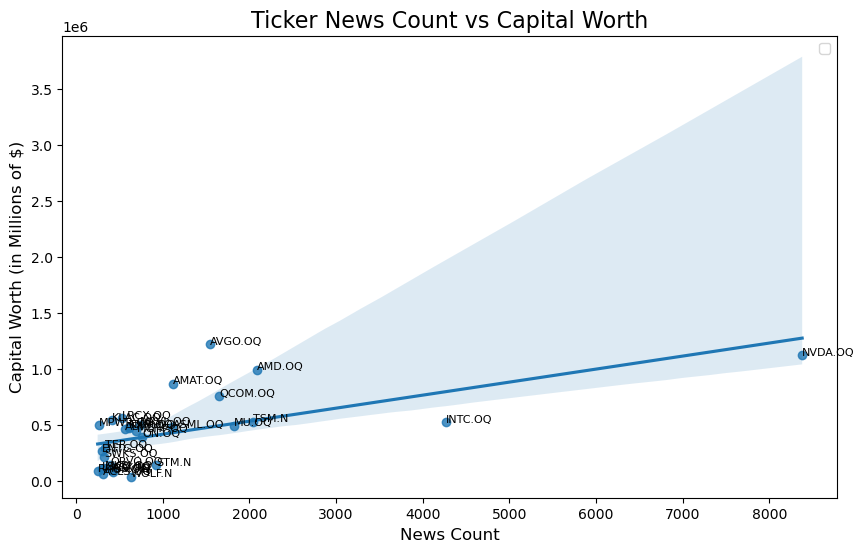}
    \caption{Actual Linear Relationship of Ticker news count weight vs Capital worth weight}
    \label{fig:countvsweight}
\end{figure}

A zoom on the left bottom side of the plot is provided by filtering out NVDA.QQ and INTC.QQ from the ticker list.

\begin{figure}[H]
    \centering
    \includegraphics[width=0.8\textwidth]{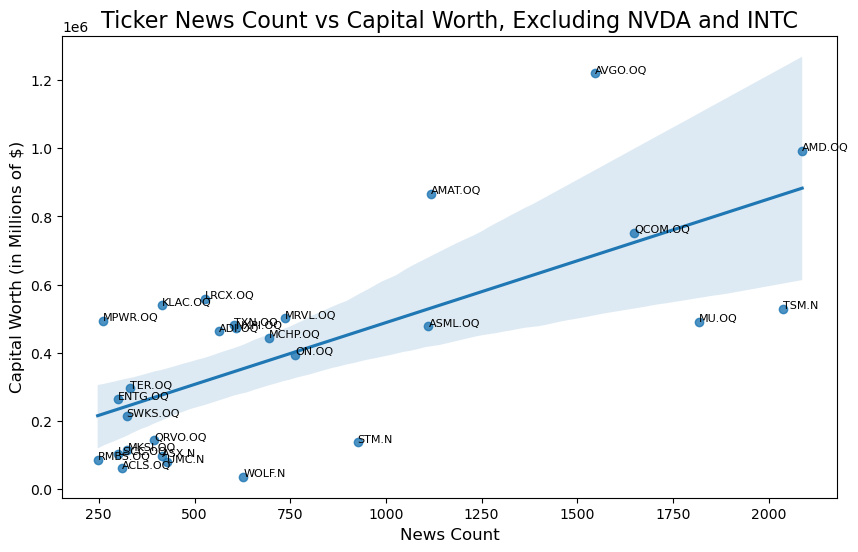}
    \caption{Actual Linear Relationship of Ticker news count weight vs Capital worth weight, Excluding NVDA and INTC}
    \label{fig:countvsweight_ex2}
\end{figure}

\begin{figure}[H]
    \centering
    \includegraphics[width=0.8\textwidth]{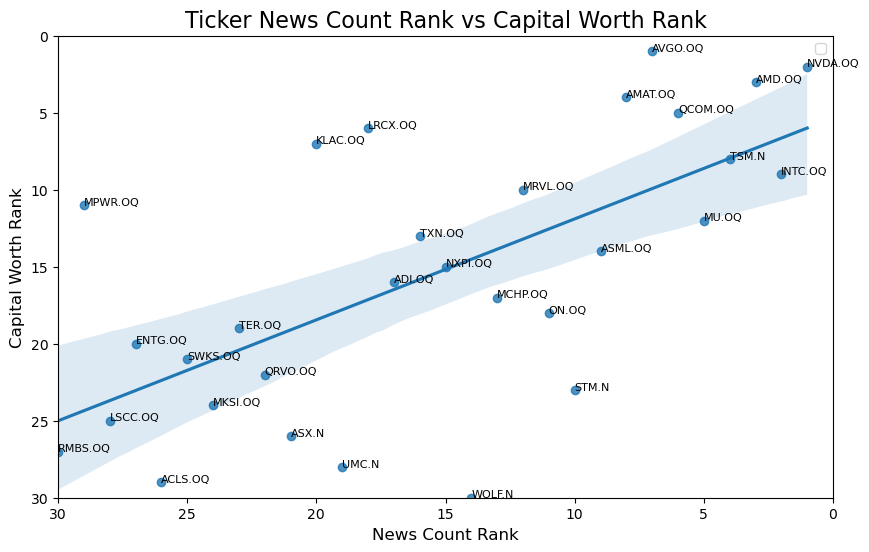}
    \caption{Ideal Linear Relationship of Ticker News Count Rank vs Capital Worth Rank}
    \label{fig:ticker_rank}
\end{figure}

We wish to provide a sophisticated method to change the relationship from the first plot in Figure \ref{fig:countvsweight} to the second in Figure \ref{fig:ticker_rank}.

\subsubsection{Memory and Weight}

Quantifying the bias of news sources and optimizing compound sentiment scores are key to improving sentiment-based forecasting models. We propose new parameters for calculating the daily compound sentiment score. Rather than using daily averages in equation \ref{eq:SentScore} as a benchmark for \( Sent_d \) (sentiment score on the date or time \( d \)), we assign weights to each component (ticker) based on its sector weight:

\begin{equation}
    \begin{aligned}
    Sent_d =  & \ \sum_{i=1}^n  \omega_{i} \, SentScore_{i, d } 
    \end{aligned}
    \label{eq:new_sentiment_equation}
\end{equation}

Here:
\begin{itemize}
    \item \( d \): represents the current day.
    \item \( n \): denotes the total number of tickers studied.
\end{itemize}

We introduce a novel approach to account for time-memory effects in market reactions. Historical news sentiment impacts future market performance, with older news exerting diminishing influence. For instance, the collective sentiment on a Monday may be shaped not only by weekend news but also by earlier events. We propose weighting algorithms to capture these decay and lagging effects, ensuring older events carry less influence but some persistence.

\begin{align} \label{eq:recursive_sentiment}
    Sent_d &= \omega_{\text{today weight}} \sum_{i=1}^n  \omega_{\text{component weight}}   SentScore_{i, d} \\
           &+ \omega_{\text{past weight}} Sent_{d-1} \notag
\end{align}

This equation includes:
\begin{itemize}
    \item Present sentiment: A weighted sum of sentiment scores adjusted for news bias and component weights (\( \omega_{\text{component weight}} \)).
    \item Past sentiment: The sentiment from the previous day (\( Sent_{d-1} \)) is included, weighted by its relevance over time.
\end{itemize}

By incorporating these weights, the model ensures that the sentiment score reflects both the immediate impact of current news and the lingering effects of past sentiment, thereby improving the accuracy of market forecasts.


The primary goal is to determine the extent to which news sources exaggerate or understate events and how this impacts market sentiment. By incorporating bias weights, memory effects, and component-specific adjustments, the model aims to better reflect true market sentiment, minimizing the effects of biased reporting. Future improvements could involve more advanced bias detection algorithms and real-time integration of data from platforms like Eikon.

Following equation \ref{eq:recursive_sentiment} and using the semiconductor sector as an example, we analyze how individual asset components contribute to the overall index. Considering 30 companies from the SOXX, we assign weights to each ticker based on their market share. For example, Nvidia (NVDA), valued at \$1,120,866.62 million and accounting for 8.64\% of the sector, has a component weight of 8.64\%.

\begin{equation}
\sum_{i=0}^{n=29} \omega_i = 1
\end{equation}

We propose that shifts in sentiment direction affect the relevance of historical data. This is incorporated into the construction of indicator functions in the new sentiment score equation, adjusting the weight of past data based on directional changes.

\subsubsection{The 3 Criteria}

We examine the impact of three types of weights on market sentiment analysis:

\textbf{1. Market Component Weights vs News Coverage:}

Certain stocks receive disproportionate media coverage compared to their actual weight in the market or sector. For instance, Nvidia may garner excessive news coverage relative to its actual weighting in the semiconductor sector.

To address this imbalance, we propose a solution that adjusts the contribution of individual tickers' sentiment to the overall sentiment of the industry or sector. This approach aims to mitigate the disproportionate influence of certain stocks, progressing from the current scenario (top figure) to a more balanced perspective (bottom figure).

Currently, the approach is to train the algorithm to learn the best weight for each individual component

\textbf{2. News Source Bias Weights:}

We recognize that news reports are rarely perfectly objective. Different news sources have biases that can affect sentiment. For example, CNN may exhibit a liberal bias and present a relatively positive outlook on green energy, whereas FOX may be more critical of gun control policies. Similarly, the BBC may adopt a more critical stance on the Chinese market compared to state-owned Chinese media, with biases shaped by the source's audience and language.

To correct for these biases, we propose assigning weights to different news sources based on the degree of bias and applying these weights when analyzing sentiment data.

To quantify the news bias, we plan to use data from different media sources and predict the results separately. We assume that a more accurate forecast comes from a more objective data source. And we will apply the learned degree of bias to assign new weights to compute the adjusted sentiments.

\textbf{3. Weights of Past News Data (Memory):}

We have already demonstrated that adjusting sentiment using indicator functions, coupled with ML techniques to optimize parameters, can enhance the accuracy of market forecasts.

Building on this, we aim to derive a new score equation that captures how the weighting of past news data (memory) influences current prices. This weighting function will help us better understand how historical sentiment data affects current market dynamics.

In summary, we focus on evaluating and optimizing the 3 criteria to enhance the sentiment score equation and NLP approach. These enhancements ensure a balanced and accurate representation of market dynamics by correcting for media attention imbalances, accounting for source biases, and incorporating memory effects. This foundational framework not only refines sentiment modeling but also provides the necessary groundwork for the development of an adjusted sentiment score equation. In the next section, we extend these concepts to introduce a more robust equation that integrates component weighting and historical memory effects for greater predictive accuracy.

\subsection{Adjusted Sentiment Score Equation}
\label{section: Adjusted Sentiment Score Equation}
This section presents an enhanced sentiment score calculated based on equation \ref{eq:recursive_sentiment} that incorporates the weighting of individual components within a portfolio and accounts for historical memory effects from past news data.


We incorporate only the most recent significant sentiment shift event. If multiple shifts have occurred but the threshold for disregarding prior data has not been met, only the most recent shift is considered, and sentiments from earlier dates are neglected.

\begin{equation}
    \begin{aligned}
     SentAll_d = & \sum_{i=1}^n \mathbbm{1}_{ Sent_d, Sent_{d-1}} \quad \omega_{\text{weight}_{i, d}}  SentScore_{i,d} \\
     = & \omega_{\text{today weight}} Sent_d\\
     &- \mathbbm{1}_{ Sent_d < 0 \ \&  \ Sent_{d-1} > 0 }  \quad \omega_{\text{past weight 1}} \,  SentAll_{d-1}\\
    &+ \mathbbm{1}_{ Sent_d > 0 \ \&  \ Sent_{d-1} < 0 } \quad \omega_{\text{past weight 2}} \,  SentAll_{d-1}\\
    &+ \mathbbm{1}_{ Sent_d  \times Sent_{d-1} \geq 0 } \quad \omega_{\text{past weight 3}} \, SentAll_{d-1}
    \end{aligned}
         \label{eq:indicator_function}
\end{equation}

We consider the influence of historical sentiment. The $SentAll_d$ variable represents the compound daily sentiment score of $n$ tickers, taking into account the cumulative influence of historical data.

We categorize sentiment changes into three cases to assess the role of historical data in the overall sentiment function, based on indicator functions:

\begin{enumerate}
    \item $\mathbbm{1}_{ Sent_d < 0 \ \&  \ Sent_{d-1} > 0 }$ : when the daily sentiment shifts from positive to negative.
    \item $\mathbbm{1}_{ Sent_d > 0 \ \&  \ Sent_{d-1} < 0 }$: when the daily sentiment shifts from negative to positive.
    \item $\mathbbm{1}_{ Sent_d  \times Sent_{d-1} \geq 0 } $ : when the daily sentiment directions stay unchanged
\end{enumerate}

When the indicator \(\mathbbm{1}_{ Sent_d < 0 \ \&  \ Sent_{d-1} > 0 }\) is met, higher volatility is likely to occur as compared to the condition \(\mathbbm{1}_{ Sent_d > 0 \ \&  \ Sent_{d-1} < 0 }.\)

 The real-world implication leads us to hypothesize that the more positive the overall sentiment is (the greater $SentAll_{d-1}$ is), the more impacts it will cause from negative news occurrences to traders' panic sales and the drop in market prices (the smaller $SentAll_d$ becomes, 
and therefore introducing the negative sign before $\mathbbm{1}_{ Sent_d < 0 \ \&  \ Sent_{d-1} > 0 }$.



We hypothesize that when sentiment shifts from positive to negative, the impact becomes more significant as the magnitude of the change increases. This would directly correlate with the weight we aim to introduce here.

Weights below $0.005$ are disregarded. For instance, after multiple recursive applications, if $\omega_{\text{time weight}, d-10} = 0.003 < 0.005$, data for dates on or before $d-10$ are excluded from consideration.

In the following section, we apply the enhanced sentiment score equation within an updated NLP framework, testing its predictive power using various algorithms and assessing its impact on market forecast accuracy.

\subsection{New NLP Approach and Forecast Results} \label{section: New NLP Approach and Forecast Results}

This section summarizes the ML methodologies, procedures, and forecast results used in the new NLP approach, which is based on the enhanced sentiment score equation.

\subsubsection{Methodology}

The primary method introduced in Deveikyte et al relied on Latent Dirichlet Allocation \cite{deveikyte2022}. In contrast, we use in this study unsupervised ML models as benchmarks to compare the performance of enhanced computational and optimization techniques.

Our objectives are twofold: first, improve the forecasting accuracy of the original model, and second, demonstrate the benefits of incorporating arithmetic computations and optimizations beyond standard LDA.

\begin{figure}[H]
    \centering
    \includegraphics[width=0.75\textwidth]{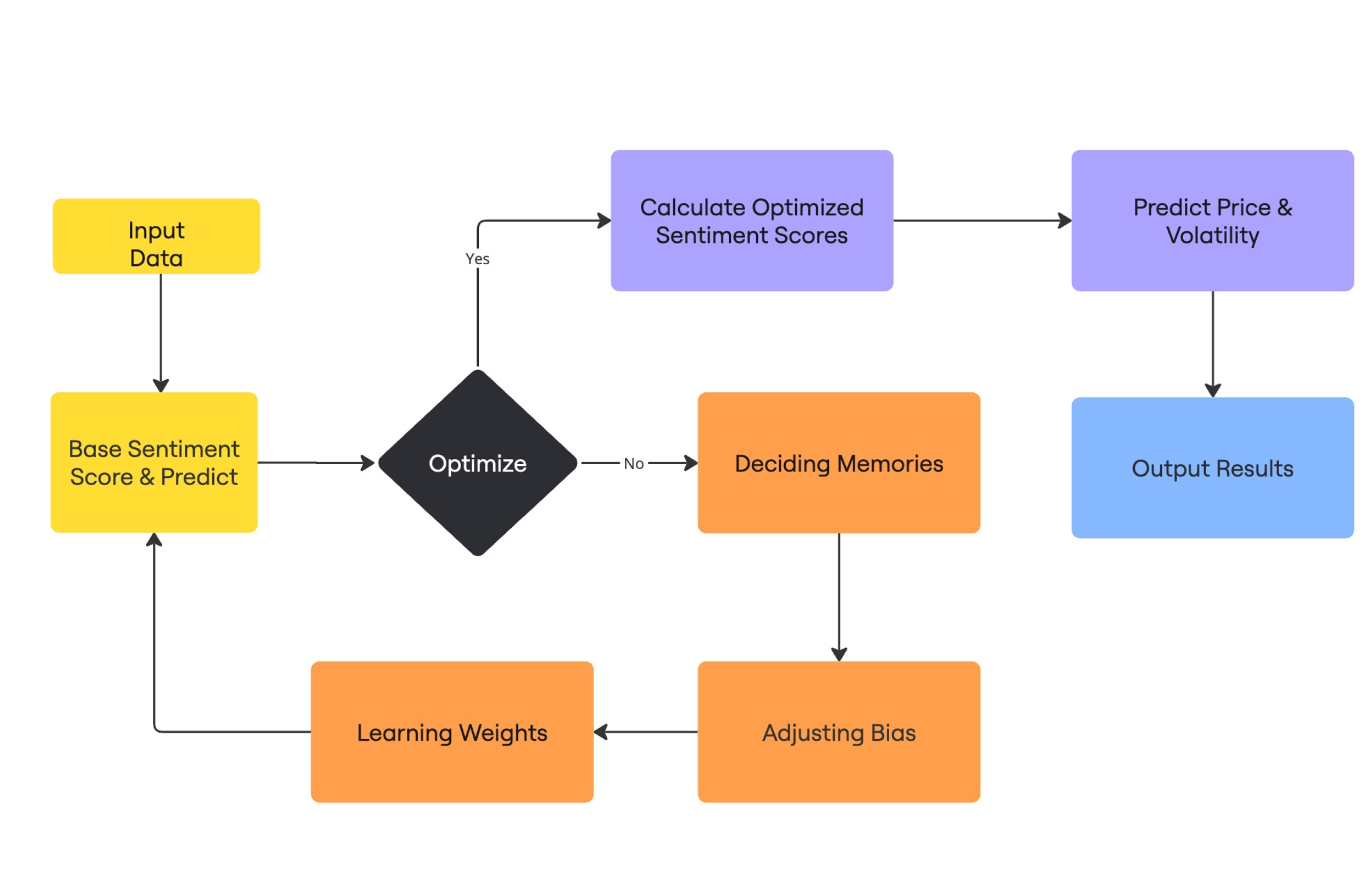}
    \caption{Algorithm Flow Chart}
    \label{fig:algorithm_flow_chart}
\end{figure}

The optimization algorithms used include models such as LDA, Logistic Regression, Ordinary Least Squares (OLS) regression. The primary focus of this paper is not to fine-tune ML models but to demonstrate the improvements gained from the new sentiment score data and procedure.

\subsubsection{Machine Learning Procedures}

We employ several ML approaches to predict stock market movements based on sentiment data. The sentiment data is first processed by converting the date index into a standardized date format. This data is then merged with stock market data, including daily market returns and volatility metrics.

The methodology involves iterating through various sentiment indicators and performing data splits into training, validation, and test sets. For each indicator, we run the model up to 1,000 times with different random states to ensure robust results.

Ordinary Least Squares (OLS) regression is used to quantify the relationship between sentiment scores (independent variable \( X \)) and daily market return or volatility (dependent variable \( Y \)). The regression provides an R-squared value to measure the proportion of variance in \( Y \) explained by \( X \). This allows us to evaluate the predictive power of sentiment scores in explaining market dynamics.

Logistic regression is employed to predict the direction of daily market returns (dependent variable \( Y \)) based on the compound sentiment score (independent variable \( X \)). Here, \( Y \) is a binary variable indicating positive or negative market movements. The accuracy of the logistic regression model serves as a key metric for assessing its performance.

The optimal models are selected based on the highest validation accuracy (for logistic regression) and R-squared values (for OLS regression) across iterations. These models are subsequently evaluated on the test dataset, with metrics including accuracy, precision, recall, and confusion matrices. The results highlight the most effective sentiment indicators for predicting market movements, offering insights into the relationship between sentiment data and market behavior.
 
Table \ref{tab:sentiment_scores_sample} shows a sample of modified sentiment scores based on different parameters in equation \ref{eq:indicator_function}.

\begin{table}[H]
\centering
\caption{Sample Sentiment Scores with Selected Parameters}
\begin{tabular}{|c|c|c|c|c|c|}
\hline
\textbf{Date} & \textbf{Base Sentiment} & \textbf{Parameters A} & \textbf{Parameters B} & \textbf{Parameters C} & \textbf{...} \\
\hline
2023-04-27 & 0.043192 & 0.043192 & 0.043192 & 0.043192 & \dots \\
2023-04-28 & 0.186383 & 0.186383 & 0.207979 & 0.229574 & \dots \\
2023-04-29 & 0.040962 & 0.040962 & 0.144951 & 0.270536 & \dots \\
2023-04-30 & -0.036497 & -0.036497 & -0.036497 & -0.036497 & \dots \\
2023-05-01 & 0.078847 & 0.078847 & 0.078847 & 0.078847 & \dots \\
\hline
\end{tabular}
\label{tab:sentiment_scores_sample}
\end{table}

Each column represents the adjusted sentiment scores based on different sets of parameters.

The model results are analyzed in Section \ref{section: Model Results}.

\newpage
\section{A Hype-Adjusted Probability Measure} 
\label{section: Hype-Adjusted Probability Measure}

In the context of financial markets, hype refers to the amplification of attention or sentiment around a particular stock, sector, or market event that exceeds its fundamental or intrinsic importance. Hype is often fueled by disproportionate media coverage, speculative behavior, or investor overreaction to news. 

A hype scenario is identified when there is a measurable and disproportionate increase in examples like:
\begin{enumerate}
    \item Media Coverage: A significant spike in the volume of news articles, social media mentions, or other sources of information about a specific stock or sector compared to its baseline or relative importance.
    \item Market Over Reactions: Corresponding anomalies in price movement and volatility, such as sharp increases or dramatic swings, that deviate from historical patterns.
    \item Imbalance in Representation: Evidence of over or under-representation in news coverage compared to the stock's weight in its sector (e.g., market capitalization).
\end{enumerate}

These indicators form the foundation for quantifying and incorporating hype into the proposed probability measure. A real-world case of Nvidia's hype is examined later in this section.

\subsection{Intuition of the New Measure} \label{section: Intuition of the New Measure}

Recall the 3 criteria:

\begin{itemize}
    \item Market Component Weights vs news Coverage"
    \item News Source Bias Weights
    \item Weights of Past News Data (Memory)
\end{itemize}

These parameters capture the influence of market components, news biases, and historical data retention (memory) on sentiment and subsequently on the market forecast.

Also, recall that we have shown from figure \ref{fig:Distribution_semtiment} that a slight negative skew of market news sentiments is observed and there exists an imbalance of news report counts vs. component weights of Nvidia among the semiconductor sector tickers. 

In addition, the news source bias weights are currently being investigated for further research. For example, in America, major news media have political leanings, CNN is more left-wing and promotes clean energy and green targets, while Republican-favored Fox argues for the opposite tone. We wish to quantify the bias instead of relying solely on polls and surveys.

To illustrate the concept of market `hype', we examine Nvidia, the most trending AI company as a case study for summer 2024.

\begin{figure}[H]
\centering
\includegraphics[width=0.8\textwidth]{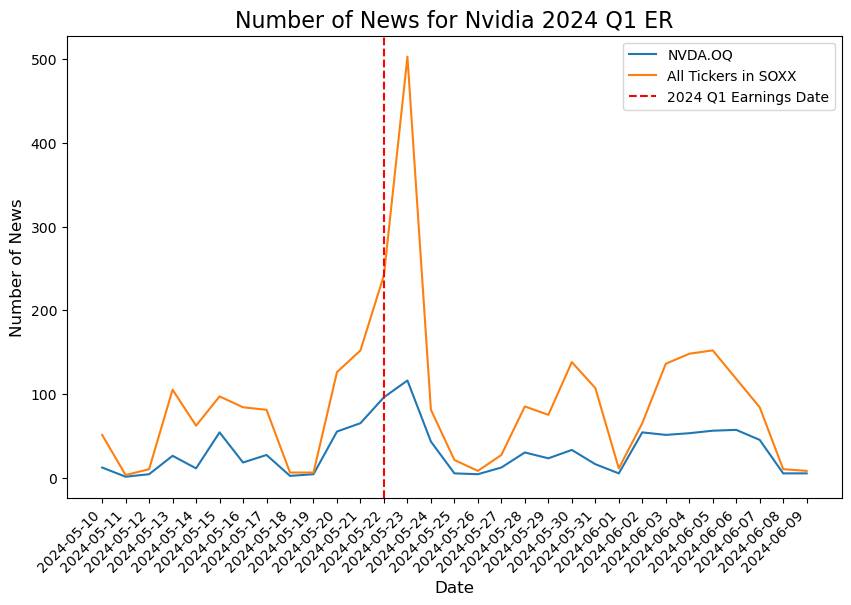}
\caption{Number of News for Nvidia Q1 Earning Report on May 22, 2024, versus tickers in SOXX}
\label{Figure: Number_News_Nvidia}
\end{figure}

In figure \ref{Figure: Number_News_Nvidia}, the blue line indicates the news counts on Nvidia from May 10 to June 9, 2024, while the red dashed line indicates May 22, 2024, when Nvidia's Q1 earnings report was released. The orange trajectory describes the total number of news collected for all tickers in the SOXX sector. We notice a rapid increase in market hype centering around the report release time.

\begin{figure}[H]
\centering
\includegraphics[width=0.8\textwidth]{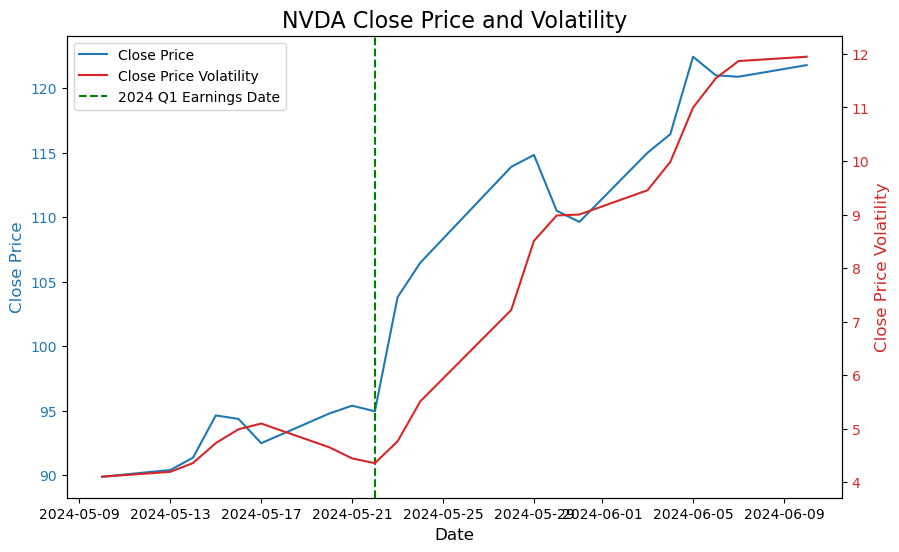}
\caption{Price and Volatility Trajectories for Nvidia Q1 Earning Report on May 22, 2024}
\label{Figure: Nvidia_ER_Price_Volatility}
\end{figure}

Figure \ref{Figure: Nvidia_ER_Price_Volatility} presents the historical movements of Price and Volatility based on the market close data for the same time frame of Figure \ref{Figure: Number_News_Nvidia}, it is on observation of a dramatic increase in both the close price and the close price volatility. 

Notice, while we do not assert a direct positive correlation between news count increases and price or volatility changes, these observations suggest that market hype can significantly impact these variables, which inspires the proposal of a hype-adjusted probability measure, \( \mathbb{P}^a \).

\subsection{Construction of A Hype-Adjusted Probability Measure}


Based on the previous intuitions, we define a hype-adjusted probability measure, \( \mathbb{P}^a \), to account for market sentiment and the existence of news biases.

\begin{definition}[Hype-adjusted Probability Measure] \textbf{Hype-adjusted Probability Measure}:
\label{def: Hype-adjusted Probability Measure}

We consider a probability space $(\Omega, \mathcal{F}, \mathbb{P})$ where $\Omega$ is the set of states of nature $\omega$, $\mathcal{F}$ is the filtration of information, and $\mathbb{P}$ the physical probability measure. We define a new probability measure \( \mathbb{P}^a \) on $(\Omega, \mathcal{F})$ by assigning new weights to the states of nature, where the effects of news weight and bias from media sources are corrected, reducing the over / under representation in the news of some particular stocks of a sector.

\end{definition}

We get the economic inspiration of change of measure from the physical measure $\mathbb{P}$ to a Harrison and Kreps (1979) ``risk-neutral" / equivalent-martingale measure; and Geman's Forward measure (1989). Note that in these two famous cases like in our setting below, the change of measure and its weights reflect the economic problem to address. 

We suppose that there are 3 states of nature $\omega_1, \omega_2,  \omega_3$ in the universe $\Omega$,  labeled as Up, Level, and Down.

We know from measure theory that the relationship between a probability measure $\mathbb{P}$ on $(\Omega, \mathcal{F})$
and an equivalent probability measure $\mathbb{P}^a $ has the following form:

\begin{equation} 
    Z = \frac{d\mathbb{P}^a}{d\mathbb{P}},
\end{equation}

\begin{equation} 
    Z(\omega) = \frac{\mathbb{P}^a(\omega)}{\mathbb{P}(\omega)}.
\end{equation}

where $Z$ is a random variable with 

\begin{equation}
   \mathbb{E}_{\mathbb{P}}[Z] = 1. 
\end{equation}

In order to define \( Z \), we need to choose \( Z(\omega_1) \), \( Z(\omega_2) \), and \( Z(\omega_3) \).

We consider now two companies, Nvidia and Intel, and their respective stock and sentiment performances:

\begin{itemize}
    \item Nvidia: Today, the stock increased by 10\%, but the sentiment increased by 20\%. Nvidia is thus considered ``overhyped", we choose: 
    \[ Z(\omega_1) < 1 \]
    \item Intel: The stock decreased by 20\%, but the sentiment only decreased by 10\%. Intel is thus considered ``underhyped", we choose: 
    \[ Z(\omega_3) > 1.\]
\end{itemize}



1. Up State (\( w_1 \)):

For Nvidia, which is overhyped, we choose \( Z(\omega_{\text{up}}) < 1 \), reducing the probability of the Up state. This reflects the fact that the market sentiment is too optimistic relative to the actual stock performance.

\[
\mathbb{P}^a(\omega_{\text{up}}) = Z(\omega_{\text{up}}) \cdot \mathbb{P}(\omega_{\text{up}}),
\]

such that

\[
\mathbb{P}^a(\omega_{\text{up}}) < \mathbb{P}(\omega_{\text{up}}).
\]

2. Down State (\( w_3 \)):

For Intel, which is underhyped, we choose \( Z(\omega_{\text{down}})> 1 \), increasing the probability of the Down state to reflect the understated sentiment relative to the actual performance:

\[
\mathbb{P}^a(\omega_{\text{down}}) = Z(\omega_{\text{down}}) \cdot \mathbb{P}(\omega_{\text{down}}),
\]

such that

\[
\mathbb{P}^a(\omega_{\text{down}}) > \mathbb{P}(\omega_{\text{down}}).
\]

3. Level State (\( w_2 \)):

In the ``Level" state, sentiment and stock price changes are neutral, and no adjustment is needed. The value we assign to \( Z(\omega_{\text{Level}})\) is such that 

\[
\mathbb{E}_{\mathbb{P}}[Z] = 1.
\]

The same construction can be extended to include more stocks.

Recalling its definition given in equation \ref{eq:conditional_expectation}, the conditional expectation under the hype-adjusted probability measure can be expressed in integral form as:

\begin{equation}
    \mathbb{E}^{\mathbb{P}^a}[X \mid \mathcal{F}] = \int_{\Omega} X(\omega) \, d\mathbb{P}^a(\omega \mid \mathcal{F}),
\end{equation}

where \( X(\omega) \) represents the sentiment score for each state \( \omega \), and the expectation is taken under the adjusted measure \( \mathbb{P}^a \).

Additionally, the expectation of \( X \) under \( \mathbb{P}^a \) can also be written in terms of the Radon-Nikodym derivative \( Z\):

\begin{equation}
    \tilde{\mathbb{E}}[X] = \mathbb{E}[X \cdot Z],
\end{equation}

where  \( \mathbb{E}[X \cdot Z] \) is the expectation under the original measure \( \mathbb{P} \), weighted by \( Z\).

Finally, the adjusted sentiment score \( X \) under \( \mathbb{P}^a \) can be rewritten in terms of its relationship with \( Z \), with \( s < t \), as:

\begin{equation}
    \mathbb{E}^{\mathbb{P}^a}[X \mid \mathcal{F}(s)] = \frac{1}{Z(s)} \mathbb{E}[X \cdot Z(t) \mid \mathcal{F}(s)].
\end{equation}

This adjustment allows us to account for market hype by converting the original news weight of a particular ticker to its correct proportion within the sector. The following section \ref{section: Model Results} presents the improved NLP forecasting results using a hype-adjusted probability measure.

\newpage
\section{Discussion}
\label{section: Model Results}

The results presented in this section highlight the effectiveness of the proposed sentiment-based framework in improving market return and volatility forecasting. By integrating advanced sentiment modeling techniques, we observe substantial accuracy improvements, which pave the way for innovative theoretical contributions and practical applications in financial analysis.

\subsection{Results}

Under a hype-adjusted probability measure of the selected date range, we calculate that the expected average difference between the original volatility and the adjusted one, based on the weights provided in the semiconductor dataset, is approximately $-0.0068$ (or $-0.68\%$). We note however, that despite this small number, the accuracy of the prediction was improved by $8\%$.

\begin{figure}[H]
    \centering
    \includegraphics[width=0.75\textwidth]{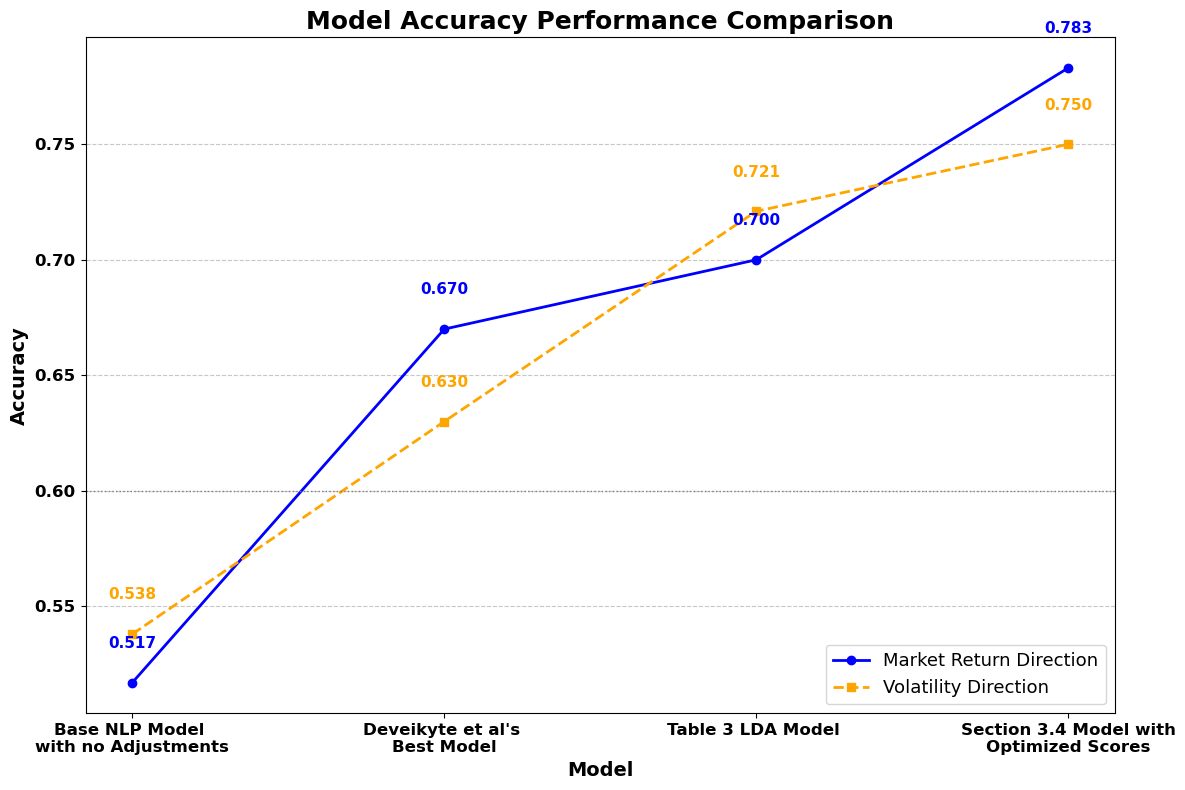}
    \caption{Model Accuracy Comparison for Return and Volatility Forecasting (discussed below)}
    \label{Figure: Model Accuracies}
\end{figure}

We choose volatility and market return directions (increase or decrease) as the target of prediction, which aligns with the observation results from sections \ref{section: New NLP Approach and Forecast Results} and \ref{section: Intuition of the New Measure}.

Figure \ref{Figure: Model Accuracies} demonstrates the progression of  accuracy across different models. In the prediction of market return direction, validation accuracy improves from $51.7\%$ in the baseline model to $70.0\%$ with a LDA model (as presented by Deveikyte et al \cite{deveikyte2022}) using adjusted scores, and further to $78.3\%$ with the optimized scores. Similarly, for volatility direction, accuracy increases from $53.8\%$ in the baseline model to $72.1\%$ and $75.0\%$ in the corresponding models. These substantial improvements validate the effectiveness of our methodology.

\subsection{Implications}

The observed accuracy increase of $+8.3\%$ for market return direction and $+2.9\%$ for volatility direction highlights the refined precision achieved through optimized sentiment modeling. This improvement translates into more reliable predictions of market trends, a critical factor for decision-making during high-volatility scenarios and market crises.

The results have demonstrated significant enhancement of the NLP forecasting approach with the adjusted scores and thus further support the validation of a hype-adjusted probability measure. 

This research extends classical finance theory by integrating sentiment adjustments into probabilistic modeling, bridging asset pricing and NLP methodologies. It formalizes sentiment as a quantitative factor, enriching models of asset returns and volatilities while addressing bias, memory, and directional shifts in market dynamics. The work represents a foundational interdisciplinary contribution, connecting finance, probability theory, and machine learning.

The model enhances forecasting accuracy for investment strategies by correcting sentiment biases, enabling better-informed decision-making. Policymakers can leverage these insights to understand sentiment-driven market behavior and regulate excessive hype or misinformation. Additionally, the framework aids in risk management and thematic investing by identifying overhyped sectors and improving resource allocation.

\subsection{Limitations and Future Research}

Note that we do not claim the uniqueness of the hype-adjusted probability measure \( \mathbb{P}^a \) we are proposing, for two sets of reasons:

1. The way we define ``hype"  can vary based on the approach used to link news to market sentiment. For instance, other researchers may define hype using different NLP techniques, sentiment scoring models, or thresholds. There is no canonical way of defining hype, and our chosen approach is just one of many possibilities to bridge news sentiment with market behavior.

2. Moreover, there could be multiple valid constructions of \( \mathbb{P}^a \) that align with the sentiment adjustments. 


The flexibility in both defining hype and constructing the adjusted measure ensures the generality of the approach, allowing it to be adapted to various datasets and market contexts.

Regarding our potential future research, one avenue could be to derive and formalize a hype-adjusted volatility and option pricing under a hype-adjusted probability measure. 
Future work can explore integrating sentiment-guided adversarial learning frameworks, such as Long Short-Term Memory (LSTM) networks and generative adversarial networks (GANs) presented, for instance, by Zhang et al in \cite{sentiment-guided2023}, to enhance the ability of the model to adapt to dynamic market conditions. Further analysis of bias or subjectivity can also be conducted using tools like the Eikon Data API, as discussed earlier in the paper.

An important point to keep in mind is that market participants do not only trade on news. Some trade on the basis of technical analysis, such as moving averages (for commodities but for equities as well); a small category of market participants, called arbitragers, trade only if they have identified a strict arbitrage opportunity; some major hedge funds try to recognize  ``statistical arbitrage" patterns. Lastly, a category of fundamental analysts uses Capital Structure and Earnings as trading signals. All these strategies interact with each other, leading to market prices whose exact formation is beyond the scope of this paper.

\newpage
\section{Conclusion} \label{Section: Conclusion}

In this paper, we present an improved NLP approach for forecasting stock return and volatility based on a hype-adjusted probability measure, \( \mathbb{P}^a \).

Besides validating improvements in financial forecasting, these results lay the foundation for broader theoretical applications. A hype-adjusted probability measure is introduced to quantify and integrate market hypes, extending the framework beyond classical sentiment analysis.

Our model evaluates text tokens with greater accuracy, effectively capturing the influence of correlations and weights across subsets. A key advantage is its low sensitivity to the accuracy of the initial token scoring method, making it robust across varied data sources. The enhanced performance of our sentiment model is exhibited in the key result figure which shows the higher accuracy obtained in predicting market responses compared to baseline models.

Lastly, our hype-adjusted probability measure can arguably be valued as a theoretical bridge between the probabilistic finance of Asset Pricing and NLP prediction in Finance, two fields which are likely to intersect more and more in the near future.


\newpage
\section{Acknowledgements}

Zheng Cao would like to express his deepest gratitude to his mentor Professor H\'{e}lyette Geman for her invaluable teaching and supervision. Her guidance has been instrumental in shaping his understanding and approach. The term ``hype-adjusted probability measure," \( \mathbb{P}^a \), which she proposed, is both brilliant and inspiring.

We also wish to acknowledge the support and advice from Professor Zhen-Qing Chen and Jason Xie of the University of Washington, Wenyu Du of Yale University, and Professor James Schmidt, Xinyu Chang, and Kai Chen of Johns Hopkins University. Additionally, we appreciate Jieruo He of Johns Hopkins University's assistance with data collection and constructing the sentiment score equation \ref{eq:indicator_function}.


\newpage
\printbibliography

\newpage
\appendix

\section{Appendix A: Ticker News Weight Table}
\label{app:ticker_weight_table}

For the table below, \textit{Close Price} refers to the price per share at market close in U.S. dollars; \textit{Capital} is reported in millions of dollars; \textit{Capital Weight \%} represents the company's market capitalization as a percentage of the entire sector's capitalization; and \textit{News Weight \%} indicates the proportion of news coverage the company receives relative to the whole sector.

Note, in the first row, the remarkable difference between columns 3 and 4.

\begin{longtable}{|l|r|r|r|r|}
\hline
\textbf{Ticker Name} & \textbf{Close Price} & \textbf{Capital} & \textbf{Capital Weight \%} & \textbf{News Weight \%} \\
\hline
NVDA.OQ & 131.38 & 1120867.0 & 8.66 & 24.52 \\
INTC.OQ & 34.59 & 524362.2 & 4.05 & 12.49 \\
AMD.OQ & 177.1 & 992494.8 & 7.67 & 6.11 \\
TSM.N & 184.52 & 528702.9 & 4.08 & 5.96 \\
MU.OQ & 131.14 & 490637.9 & 3.79 & 5.32 \\
QCOM.OQ & 207.12 & 752063.7 & 5.81 & 4.82 \\
AVGO.OQ & 1733.31 & 1220293.0 & 9.43 & 4.52 \\
AMAT.OQ & 251.47 & 866274.3 & 6.69 & 3.27 \\
ASML.OQ & 1059.97 & 477667.0 & 3.69 & 3.25 \\
STM.N & 41.51 & 137307.8 & 1.06 & 2.71 \\
ON.OQ & 73.48 & 394591.1 & 3.05 & 2.23 \\
MRVL.OQ & 73.84 & 501852.5 & 3.88 & 2.15 \\
MCHP.OQ & 92.34 & 444145.9 & 3.43 & 2.03 \\
WOLF.N & 23.05 & 35874.98 & 0.28 & 1.83 \\
NXPI.OQ & 274.91 & 472496.2 & 3.65 & 1.78 \\
TXN.OQ & 200.16 & 480016.7 & 3.71 & 1.77 \\
ADI.OQ & 232.01 & 462719.8 & 3.57 & 1.65 \\
LRCX.OQ & 1112.55 & 558002.5 & 4.31 & 1.54 \\
UMC.N & 8.59 & 78866.05 & 0.61 & 1.25 \\
KLAC.OQ & 874.9 & 538704.0 & 4.16 & 1.22 \\
ASX.N & 11.96 & 96245.99 & 0.74 & 1.21 \\
QRVO.OQ & 119.69 & 142502.9 & 1.10 & 1.15 \\
TER.OQ & 153.48 & 295777.2 & 2.28 & 0.97 \\
MKSI.OQ & 135.06 & 113224.3 & 0.87 & 0.95 \\
SWKS.OQ & 106.41 & 213374.7 & 1.65 & 0.94 \\
ACLS.OQ & 151.06 & 60574.45 & 0.47 & 0.91 \\
ENTG.OQ & 140.23 & 263961.8 & 2.04 & 0.88 \\
LSCC.OQ & 59.75 & 101910.3 & 0.79 & 0.87 \\
MPWR.OQ & 846.2 & 493544.3 & 3.81 & 0.76 \\
RMBS.OQ & 63.88 & 85799.32 & 0.66 & 0.72 \\
\hline
\end{longtable}

Note, this table is collected from LSEG, on July 17, 2024\cite{LSEG}.
This ticker weight table is adjusted by removing 2 values:
CME E-MINI S\&P500-TECHNOLOGY SECTOR INDEX FUTURE SEP 2024, $ 0.15247 \%$ and CME INDEX and OPTIONS MARKET E-MINI RUSSELL 2000 INDEX FUTURE SEP 2024, $0.15247 \%$.

\end{document}